\documentclass[%
reprint,
amsmath,amssymb,
aps,
]{revtex4-1}

\usepackage{graphicx}
\usepackage{dcolumn}
\usepackage{bm}

\begin{document}
	
	
	\title{Transit Ramsey EIT resonances in a Rb vacuum cell}
	
	\author{Ravn M. Jenkins}
	\affiliation{College of William \& Mary, Williamsburg, VA. 23185, USA}

	\author{Eugeniy E. Mikhailov}
	\email[]{eemikh@wm.edu}
	\affiliation{College of William \& Mary, Williamsburg, VA. 23185, USA}
	
	\author{Irina Novikova}
	\email[ixnovi@wm.edu]{}
	\affiliation{College of William \& Mary, Williamsburg, VA. 23185, USA}

	\date{\today}
	
	\begin{abstract}
		We investigate a dual-channel arrangement for electromagnetically-induced transparency in a vacuum Rb vapor cell, and report the observation of a transient spectral feature due to the atoms traversing both beams while preserving their ground-state spin coherence. Despite a relatively small fraction of atoms participating in this process, their contribution to the overall lineshape is not negligible. By adjusting the path difference between the two optical beams, the differential intensity measurement can produce an error signal for the microwave frequency stabilization as strong as a single-channel measurement, but it provides a much higher signal-to-noise ratio due to the cancellation of intensity noise, dominating the signal channel detection. 
	\end{abstract}
	
	\pacs{}
	
	\maketitle
	
	\section{Introduction}
	
	Two-photon Raman resonances in Rb atoms provide an all-optical access to long-lived spin coherences and thus enables observation in spectrally narrow optical transmission and absorption resonances, widely used for atomic clocks~\cite{vanier_book,vanier05apb,Shah201021}, magnetometers~\cite{budker_optmagn_book,stahler01eurlett,romalis2007natphys}, and other metrological applications. Since the intrinsic spin decoherence rate is negligible, the spectral width of such features is typically limited by the environmental disturbances, such as dephasing collisions or external field gradients. For the thermal atomic ensembles, the limited interaction time of moving atoms with the laser beam becomes a leading limiting factor~\cite{novikovaLPR12}, as the consequent depolarizing wall collisions completely dephase any light-induced spin coherence, unless a special anti-relaxation coating is applied~\cite{budker'98,klein2011pra_eit_coated_cells}. Often an inert buffer gas is introduced to increase the interaction time as atoms slowly diffuse through the laser beam~\cite{XiaoMPL09}, and this approach has been successfully used to reduce the electromagnetically induced transparency (EIT) resonance linewidth down to a few tens of Hz~\cite{wynands'99}. However, collisions with the buffer gas also perturb the transitions, increasing the environmental sensitivity. Plus, some two-photon resonances are degraded by the strong collisional depolarization of the optical excited state, thus making the use of buffer gas impractical. 
	
	\begin{figure}[h!]
		\includegraphics[width=1.0\columnwidth]{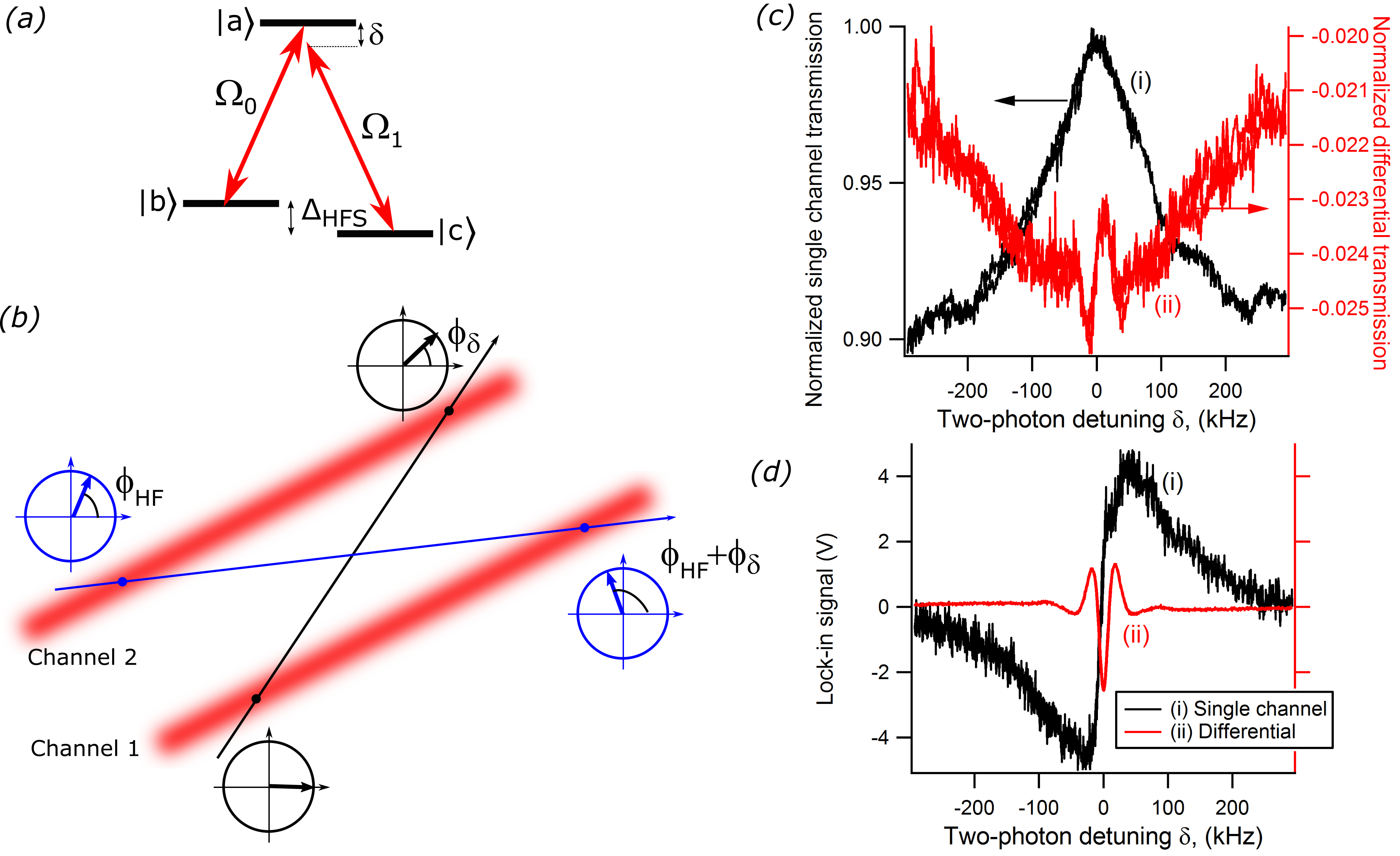}%
		\caption{(a) Schematic EIT level diagram. (b) Simplified geometry of the two-channel transient EIT setup.  The arrows in the circles indicate the dark state phases of two atoms traveling symmetrically between the beams. For this illustration we set the phase between the two EIT optical fields to be zero in the first beam and $\phi_\mathrm{HF} \ne 0$ for the second beam. In case of the non-zero two-photon detuning $\delta$, the dark state phases of both atoms evolve by $\phi_\delta = \delta\cdot \tau$ after $\tau$ transit time between the two beams, resulting in the difference in the optical response during the repeated interrogation. (c) Examples of the optical transmission  for a single-channel EIT (i) and for the intensity difference between the two channels (ii). Both signals are normalized to the peak EIT transmission in one channel Small non-zero background in the differential signal is due to imperfect match of the laser beam diameters in the two channels. (d) Same signals recorded using the phase-sensitive lock-in detection.  \label{fig:intro}}
	\end{figure}
	
	Here we explore the ballistic motion of Rb atoms between two identical illuminated regions in a Rb vapor cell with no anti-relaxation wall coating and no additional buffer gas to produce an additional narrow structure within a usual EIT resonance, as shown in Fig.~\ref{fig:intro}. Each beam consists of two optical fields near the two-photon EIT resonant conditions, shown in Fig.~\ref{fig:intro}(a). While the majority of atoms interact only once with one of the laser beams before hitting the cell wall, there is a group of atoms, depicted in Fig.~\ref{fig:intro}(b), that traverses both beams without decohering. For these atoms the conditions of the original Ramsey experiment~\cite{RamseyPhysRev.78.695} are recreated, as they experience two consecutive interactions, separated by the free evolution region. Such an arrangement is also a spatial equivalent of Raman-Ramsey CPT experiments~\cite{ZanonPRL05,ClaironIEEE09,Liu:13}, as the spin superposition state, prepared in the first beam, is allowed to evolve in the dark before interacting with the second beam.  The spectral narrowing associated with such multi-zone interactions has been investigated in the case of degenerate Hanle magneto-optical resonances~\cite{LezamaPhysRevA.81.023801,Radojicic:15}.  With the use of a bichromatic optical field to produce a coherent superposition of two ground-state levels, we observe an additional interference-like feature on top of the regular single-channel EIT resonance, shown in Fig.~\ref{fig:intro}(c,d). While the signal can be clearly observed in direct optical transmission, it is more convenient to analyze it using a phase-sensitive lock-in detection with improved signal-to-noise ratio (SNR). 
	
	This transit Ramsey EIT (TREIT) effect provides an additional way of reducing the width of the EIT feature in the situation when addition of buffer gas or antirelaxation coating is impossible, for example in ultra thin vapor cells~\cite{Sargsyan2011,SargsyanJMO2015}. It is also somewhat surprising that even though only a small fraction of atoms traverse both laser beams, their contribution to the resonance slope is comparable with the single-channel resonance amplitude. Moreover, the possibility of such spin coherence transfer may need to be accounted for in multiplexed experimental arrangements, commonly used in quantum optics~\cite{EnglandJPB2012,Higginbottom:15,FigueroaPhysRevApplied.8.034023}. 
	
	\section{Experimental arrangements}
	
\begin{figure}[h!]
		\includegraphics[width=01\columnwidth]{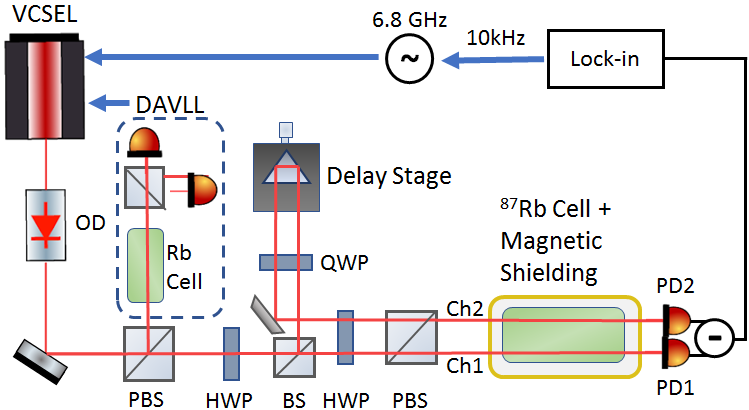} \\
		\caption{Schematic of the experimental setup for the differential detection. For a single-channel measurements one of the beams is blocked before the cell. See text for abbreviations. \label{fig:Setup}}
	\end{figure}

	The schematic of the experimental setup is shown in Fig.~\ref{fig:Setup}. To create the two optical fields for  the EIT resonance, we used a vertical cavity surface-emitting diode laser (VCSEL) operating at the Rb D$_1$ line (794.7~nm), current-modulated at the frequency of the ${}^{87}$Rb ground state hyperfine splitting $\Delta_\mathrm{HFS}\simeq 6.834$~GHz. We locked the carrier frequency of the laser to the $5S_{1/2}F=2\rightarrow 5P_{1/2}F'=1$ optical transition with dichroic atomic vapor laser lock (DAVLL), so that the +1 modulation sideband, containing $\approx 20$\% of the total laser power, became resonant with the $5S_{1/2}F=1\rightarrow 5P_{1/2}F'=1$ transition, forming a $\Lambda$ system, shown in Fig.~\ref{fig:intro}(a). A detailed description of the VCSEL current modulation and DAVLL arrangements can be found in Ref.~\cite{mikhailov2010JOSAB_linparlin_clock}. 
	
	After passing through the optical diode (OD), the VCSEL output with a maximum total power of 300~$\mu$W was  split into two beams using a non-polarizing beam splitter (BS). While the transmitted beam (Ch1) passed directly forward, the reflected beam (Ch2) was directed toward a delay line, consisting of a retroreflecting  prism mounted on a translation stage, before being reflected by a mirror to travel parallel to the first beam at a separation of $\approx 5$~mm. Moving the prism allowed us to adjust the relative phase between the zeroth and the first modulation sidebands inside the Rb cell in the second channel:
	\begin{equation} \label{RFphase}
	\phi_\mathrm{HF}=2\pi\Delta z/\lambda_\mathrm{RF},
	\end{equation}
	where $\Delta z$ is the additional path length in the delay stage, and $\lambda_\mathrm{RF}=c/\Delta_\mathrm{HFS}$ is the wavelength of the resonant frequency between the two hyperfine states. For this experiment we used the lin$||$lin EIT configuration~\cite{ZibrovPhysRevA.81.013833.2010,mikhailov2010JOSAB_linparlin_clock}, and a polarizing beam splitter (PBS) placed before the Rb cell ensured the identical linear polarization of all optical fields in both channels. Half-wave plates (HWP) and a quarter-wave plate (QWP) before and after the non-polarizing beam splitter allowed us to precisely balance the laser power in two channels. At the cell location, both laser beams had almost identical slightly elliptical Gaussian profiles: the measured $1/e^2$ radii in the first channel were $0.72$~mm and $0.75$~mm, and in the second channel $0.74$~mm and $.72$~mm. 
	
	Both beams then passed through an evacuated cylindrical Pyrex cell (length 75mm, diameter 22mm) containing isotopically enriched $^{87}$Rb vapor, heated to $44.5^\circ$~C. Then, the transmitted light intensities in both channels were detected using two identical photodiodes, PD1 and PD2, that can be operated in the differential mode. We have also recorded the output of the lock-in amplifier by superimposing an additional 10~kHz frequency modulation on the $6.834$~GHz VCSEL RF modulation signal.  
	
	\section{Transient Ramsey resonance observation}
	
	To understand the TREIT lineshape one needs to consider the relative phase between the two optical fields, forming a two-photon EIT resonance (in our case, the carrier and the first modulation sidebands of the VCSEL laser, as shown in Fig.~\ref{fig:intro}(a)). While locally the two optical fields are nearly perfectly phase coherent, the value of their relative phase changes as the beams propagate thanks to their frequency mismatch, at a rate given by Eq.(\ref{RFphase}). Thus, if the two beams travel unequal paths before entering the cell, the exact expressions for the atomic dark state in each beam will reflect the acquired phase difference, as illustrated in Fig.~\ref{fig:intro}(b). For example, if we set the relative phase between two EIT fields as zero in the first beam and $\Delta \phi_\mathrm{HF}$ in the second beam, we can write the expressions for the unperturbed EIT dark states independently formed in each channel: 
	\begin{eqnarray} \label{DS0}
	|D_1\rangle(t=0) &=& (\Omega_1 |b\rangle -\Omega_0 |c\rangle)/\Omega,  \\
	|D_2\rangle(t=0) &=& (\Omega_1 |b\rangle -e^{i \phi_\mathrm{HF}} \Omega_0 |c\rangle)/\Omega, \nonumber
	\end{eqnarray}
	where $\Omega_0$ and $\Omega_1$ are the absolute values of Rabi frequencies for the two EIT transitions, and $\Omega=\sqrt{\Omega_0^2+\Omega_1^2}$ is the normalization coefficient. In the case of zero two-photon detuning $\delta = 0$ the optical response of the atoms, prepared in the dark state in one beam and then probed by another, is symmetric for both beams. However, a small two-photon detuning $\delta$ breaks this symmetry, since during the transit time $\tau$ between the two interactions the relative phases of the both dark states evolve by the same amount $\delta\cdot\tau$:
	\begin{eqnarray} \label{DStau}
	|D_1\rangle(t=\tau) &=& (\Omega_1 |b\rangle -e^{i \delta\cdot \tau}\Omega_0 |c\rangle)/\Omega,  \\
	|D_2\rangle(t=\tau) &=& (\Omega_1 |b\rangle -e^{i \phi_\mathrm{HF}+ i \delta\cdot \tau} \Omega_0 |c\rangle)/\Omega, \nonumber
	\end{eqnarray}
	causing the difference in optical responses depending on with which beam the atoms first interacted.
	
\begin{figure}[h!]
		\includegraphics[width=0.8\columnwidth]{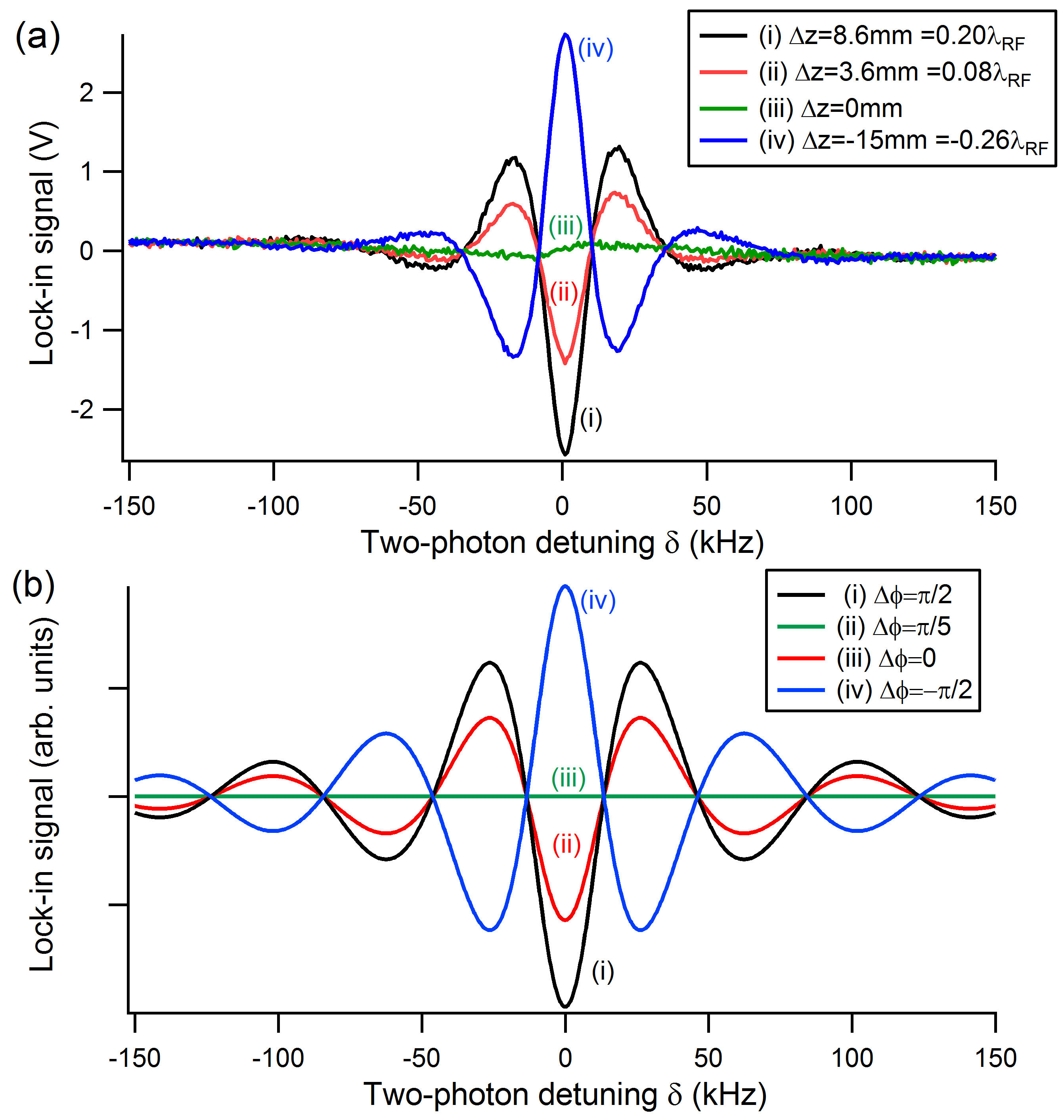}
		\caption{(a) The differential lock-in signals as a function of the two-photon detuning for different relative prism position. Laser power in each channel is $\approx 50~\mu$W. (b) Theoretical simulations of the lock-in readout of the differential TREIT signal using Eq.(\ref{TREITsig}). \label{fig:DiffPrismPosition}}
	\end{figure}

	We can experimentally verify the significance of the EIT phase difference between the two beams by controlling the beam path for the second channel using a delay stage. Fig.~\ref{fig:DiffPrismPosition}(a) shows the differential lock-in signal for the different delays. It is easy to see that it is possible to adjust the relative delay to almost perfectly match the EIT resonances in each channel (the small residual signal is due to small laser beam disbalance in two channels). However, by changing the delay one can maximize the contrast of the TREIT signal between the two channels.
	
	We can qualitatively confirm such behavior using a repeated interaction model developed in Refs.\cite{xiaoPRL06,XiaOE08} to describe the optical response of atoms after two consecutive interactions under EIT conditions. Generalizing this expression for the two regions with identical absolute values but different in phase EIT fields, we can describe the differential signal for a given travel time between the two beams $\tau$ as:
	\begin{eqnarray}\label{TREITsig}
	\Delta I(\delta)&\propto&\frac{|\Omega|^2}{\delta^2+\Gamma^2}e^{-2\Gamma t_\mathrm{tr}}\sin{\phi_\mathrm{HF}}\times \\ \nonumber
	&\times&\sin{ [ \delta(2t_\mathrm{tr}+\tau)+\tan^{-1}(\delta/\Gamma)]},
	\end{eqnarray}
	where $\Gamma$ is the power-broadened single-channel EIT linewidth and $t_\mathrm{tr}$ is the transit time of an atom through the interaction region. Here we neglect the intrinsic ground-state decoherence rate. 
	
	Fig.~\ref{fig:DiffPrismPosition}(b) shows the results of numerical simulations for the differential lock-in signal, averaged over the transverse Maxwell-Boltzmann velocity distribution of atoms, moving between the two beams. Despite many simplifications of the model (optically-thin medium, absence of Doppler broadening of optical transitions and the longitudinal motion of the atoms), the general features of the model predictions nicely match the experiment: we observe the maximum lock-in differential signal at zero two-photon detuning for $\phi_\mathrm{HF} = \pm \pi/2$, and the TREIT feature disappears for $\phi_\mathrm{HF}=0$.
	
	\section{Signal-to-noise analysis}
	
	Many EIT-based measurements suffer from the residual intensity noise, especially if broad-band lasers, such as VCSELs, are used to excite optical transitions~\cite{CamparoPhysRevA.59.728}. Several differential EIT schemes, e.g., based on magneto-optical rotation~\cite{SihongGu2015,SunOE16,Guidry:17,GuEPL2017} or polarization selection rules~\cite{kitchinIEEE2008}, have been proposed recently to suppress the common-mode intensity noise while maintaining high-contrast EIT resonant features. 
	
\begin{figure}[h!]
		\includegraphics[width=0.8\columnwidth]{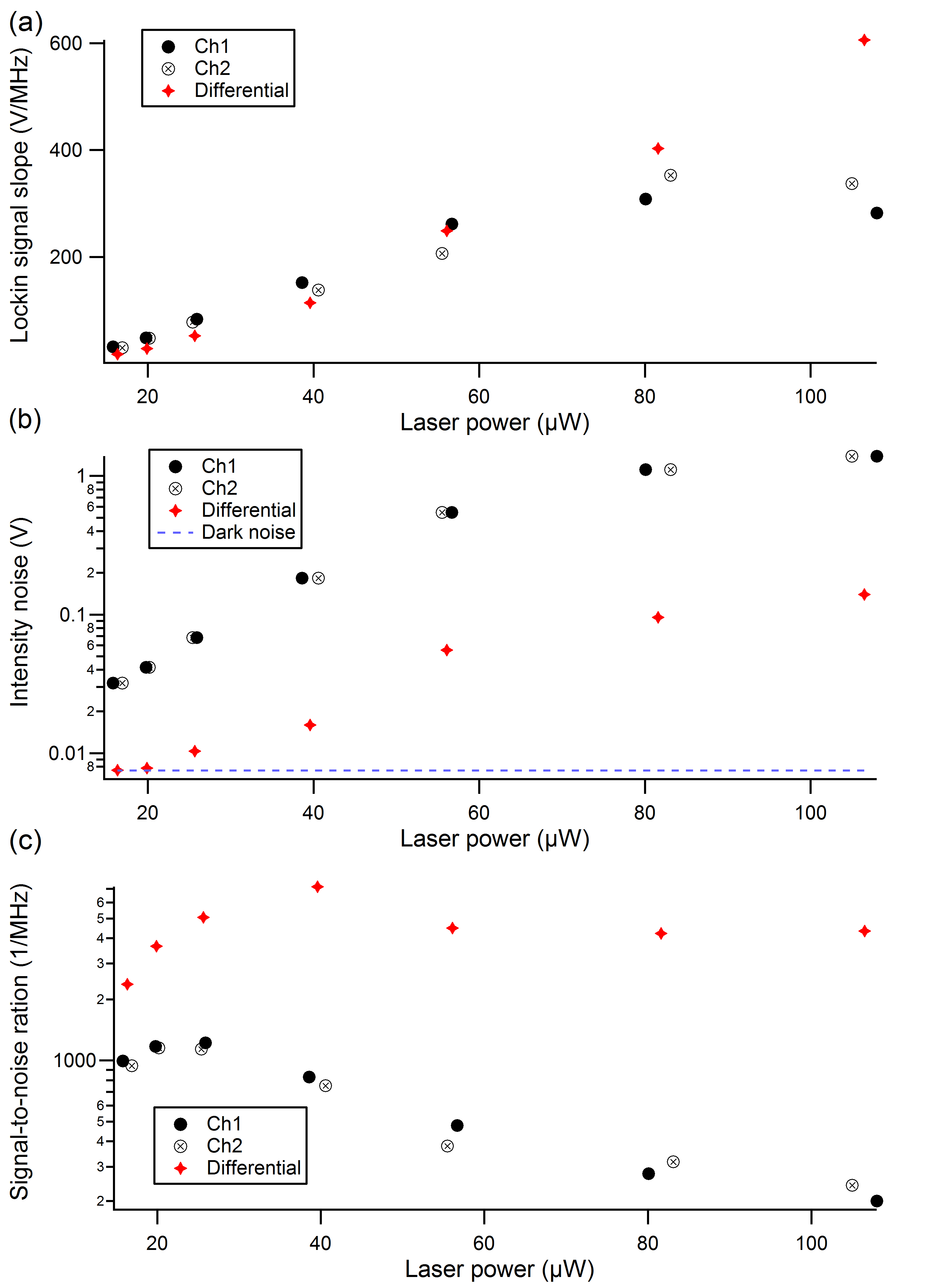} %
		\caption{Comparison of the one-channel EIT and TREIT performance. (a) Slope of the error lock-in signal for each optical channel and for the differential signal at the corresponding zero-crossing detunings. (b) lock-in noise measured at zero-crossing two-photon position. Horizontal line shows the dark electronic noise level. (c) Signal-to-noise ratio (defined as slope of the error signal divided by the measure noise). For the differential measurements the average power between the two channel is used. \label{fig:SNR}}
	\end{figure}

	Our observations suggest that the TREIT signal may also be applied to reduce the intensity noise and thus boost the signal to noise ratio. Simple visual comparison of a single-channel EIT and TREIT signals in Fig.~\ref{fig:intro}(d) shows a strong noise suppression in the differential signal. To quantify this observation we measured the slope of each lock-in signal near zero two-photon detuning at the point where it crosses zero, as this slope determines the strength of the feedback error signal if the resonance is used for frequency stabilization. The measured slope as a function of laser power is plotted in Fig.~\ref{fig:SNR}(a). It is easy to see that both EIT schemes give comparable results. A single-channel EIT performs better at lower laser power, likely due to the reduced power broadening. Higher laser power, however, improves the optical pumping of atoms into the dark state, thus increasing the number of atoms contributing to the two-beam interactions. In addition, the Ramsey interrogation schemes are known to be less sensitive to the power broadening during the evolution in the dark period~\cite{ZanonPRL05,ClaironIEEE09}. Thus, not surprisingly, the measured slope of the TREIT signal surpasses that of the single-channel EIT at higher laser powers.  
	
	At the same time, the comparison of the measured noise levels, shown in Fig.~\ref{fig:SNR}(b), clearly demonstrate the advantage of the differential detection, as we see more than an order of magnitude noise reduction for the differential TREIT signal. In fact, we were unable to accurately measure the TREIT noise for the lowest laser powers, as it fell below the technical noises of our detector. 
	
	The resulting signal-to-noise ratio (SNR) for both schemes is shown in Fig.~\ref{fig:SNR}(c). Note, that while the SNR for the single-channel EIT has a clear maximum at approximately $30~\mu$W of laser power due to the known saturation of the EIT amplitude at higher powers~\cite{mikhailov2010JOSAB_linparlin_clock}, the TREIT SNR remains relatively constant at high laser powers, which may be an attractive feature for some applications.

	\section{Conclusions}
	
	We have demonstrated a possibility to obtain a narrow differential signal on top of a regular EIT resonance taking advantage of the ballistic motion of coherently prepared atoms between two spatially separated identical optical channels in a vacuum Rb vapor cell. This feature is based on the consecutive Ramsey-like repeated interactions of atoms as they fly through both beams. We demonstrated that it is possible either to cancel or to produce a strong differential optical signal by controlling the relative phase between the two EIT fields in two regions, i.e setting the phases of the EIT dark states in the two channels. Moreover, we demonstrate that such differential detection may offer significant advantages for metrology, as it allows the subtraction of common intensity noise without sacrificing the strength of the feedback error signal.   
	
	\section{Acknowledgments}
	We would like to thank Kelly Roman for the original exploratory work with two-beam EIT configuration, and Hana Warner and Kangning Yang for their help with the initial experiment construction. This research was supported by the National Science Foundation grant PHY-308281. RMJ acknowledges the support of the DeWilde research fellowship.

%
\end{document}